\documentclass[epj]{svjour}
\usepackage{latexsym}
\usepackage{graphics}
\usepackage{epsfig}
\usepackage[usenames]{color} 

\begin{document}
\title{A model for $A=3$ antinuclei production in proton-nucleus collisions}
\author{
R.P. Duperray\thanks{duperray@isn.in2p3.fr},
K.V. Protasov\thanks{protasov@isn.in2p3.fr},
L. Derome\thanks{derome@isn.in2p3.fr}, 
and M. Bu\'enerd\thanks{buenerd@isn.in2p3.fr}
}
\institute{Institut des Sciences Nucl\'eaires, IN2P3-CNRS, UJFG,
53, Avenue des Martyrs, F-38026 Grenoble Cedex, France}
\date{Received: date / Revised version: date}
\abstract{A simple coalescence model based on the same diagrammatic approach of
antimatter production in hadronic collisions as used previously for antideuterons 
is used here for the hadroproduction of mass 3 antinuclei. It is shown that 
the model is able to reproduce the existing experimental data on $\overline{t}$ and 
$\overline{^3\mbox{He}}$ production without any additional parameter.
\PACS{{24.10.-i}{Nuclear-reaction models and methods}}}
\maketitle

%!!!!!!!!!!!!!!!!!!!!!!!!!!!!!!!!!!!!!!!!!!!!!!!!!!!!!!!!!!!!!!!!!!!!!!!!!!!!!!!!!!!!!!!!!!!!!!!!!!!!!!!!!!!!!!!!!!!!!!

\section{Introduction}

The increasing interest in the study of production of light antinuclei in proton-proton and 
proton-nucleus collisions is motivated by the presence of anti-nuclei in cosmic rays which 
has potentially important implications on the matter-antimatter asymmetry of the universe. 
From this point of view, it is important to determine the amount of anti-matter which can be 
produced in the galaxy through the interaction of high-energy protons with the interstellar 
gas. A new generation of experiments (AMS \cite{AMS}, PAMELA \cite{AD02}) should be able
to measure the flux of anti-matter in a near future.

The calculations of the $\overline{t}$ and $\overline{^3\mbox{He}}$ production cross sections 
reported here are based on the same diagrammatic approach to the coalescence model as
used recently \cite{DU03a} to describe the $\overline{d}$ production in proton-proton and 
proton-nucleus collisions.

The coalescence model \cite{CLM} is based on the simple hypothesis that the nucleons, 
produced during the collision of a beam and a target, fuse into light nuclei whenever the
momentum of their relative motion is smaller than a coalescence radius $p_0$ in the momentum space, which is 
a free parameter of the model, usually fit to the experimental data (see \cite{COALEXP} for
example). A simple diagrammatic approach to the coalescence model developed in \cite{KO92} 
provided a microscopic basis to the model. In this approach, the parameter $p_0$ is 
expressed in terms of the slope parameter of the inclusive nucleon production spectrum and 
of the wave function of the produced nucleus.

This diagrammatic approach has been generalized in \cite{DU03a} to antideuteron 
production by taking into account threshold effects and the anisotropy of the angular 
distributions. This approach can reproduce most existing data without any additional 
parameter in energy domains where the inclusive antiproton production cross sections are 
well known.

This article reports on the application of this approach to the production of $A=3$ 
antinuclei. It is the first microscopic calculation of this cross section to the knowledge 
of the authors. In \cite{CH97}, the $\overline{^{3}\mbox{He}}$ production cross section 
in proton-proton collisions was calculated using the standard coalescence model, with the 
parameter $p_0$ taken from the $\overline{d}$ production data.

Unfortunately, the experimental data required to be compare to the calculations are limited. 
Only two sets of experiments have measured the production of mass 3 antinuclei in 
proton-nucleus collisions. $\overline{t}$ and $\overline{^3\mbox{He}}$ were discovered 
at IHEP (Serpukhov), with one experimental points measured for $\overline{t}$  and one for 
$\overline{^3\mbox{He}}$ \cite{AN71,VI74}, while in the CERN experiment (SPS, WA 33) 
\cite{BO78,BU80}, four experimental points were measured for $\overline{t}$ and eight for 
$\overline{^3\mbox{He}}$.
For these latter data however, the $\overline{t}$ and $\overline{^3\mbox{He}}$ production cross 
sections were measured with respect to the pion production cross section at the same 
momentum. This requires the corresponding 
experimental values of the pion production cross section to be known to extract
the values of the $\overline{t}$  and $\overline{^3\mbox{He}}$
production cross sections. 

The article is organized as follows. The main ideas of the theoretical approach are described
in section 2. The formalism is generalized to the case of $A=3$ antinuclei production in section 
3. Section 4 is devoted to the results and the comparison to the experimental data. A brief 
summary of the work is provided before the work is concluded in the last section.

%!!!!!!!!!!!!!!!!!!!!!!!!!!!!!!!!!!!!!!!!!!!!!!!!!!!!!!!!!!!!!!!!!!!!!!!!!!!!!!!!!!!!!!!!!!!!!!!!!!!!!!!!!!!!!!!!!!!!!!

\section{Diagrammatic approach to the coalescence model}

\begin{figure}[h!]
\epsfxsize=8cm \centerline{\epsfbox{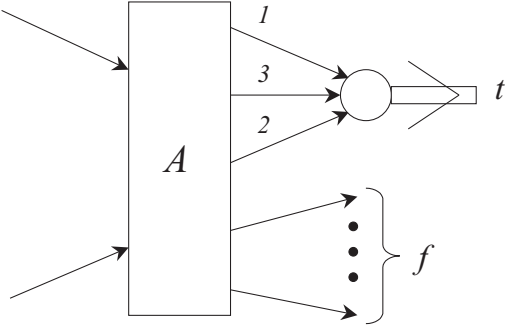}} \caption{The
simplest Feynman diagram corresponding to coalescence of three
nucleons into a tritium or $^3$He. \label{FEYN}}
\end{figure}

The main ideas of the diagrammatic approach of the coalescence model for nuclear fragment
production are reminded here for the reader's convenience \cite{KO92}. The simplest Feynman 
diagram of Fig.~\ref{FEYN} corresponding to fusion of three nucleons is considered as a basis for the 
coalescence model. Here the symbol $f$ designates the state of all particles but nucleons 1, 2 
and 3 which form the tritium or the helium 3 nucleus produced in the final state (specified by the
$t$ symbol on the graph).

%----------------------------------Fig. 1-----------------------------------------------
%---------------------------------------------------------------------------------------

The physical picture behind this diagram is quite simple: the nucleons produced in a collision 
(block A) are slightly virtual and can fuse without any further interaction with the 
nuclear field. This diagram is not the only possible contribution to the full transition 
amplitude. However, mutual cancellations of a number of other contributing diagrams result
in the diagram of Fig.~\ref{FEYN} being dominant \cite{BR82}. This diagram can be calculated using the 
technique developed in \cite{SH68}.

The probability for three nucleon coalescence is given by:
%----------------------------------------------------------------------------------------
\begin{eqnarray}
d^{3}W_{t}=\left| M\right| ^{2}\frac{m_{t}}{E_{t}}\frac{d^{3}p_{t}}{\left(
2\pi \right) ^{3}},\label{eq1} 
\end{eqnarray}
%----------------------------------------------------------------------------------------
where $m_{t}$ in the mass of the $t$ fragment and $E_{t}$ its energy in the  
(beam)nucleon-(target)nucleon center of mass system of the colliding nuclei. The probability for 
three nucleon production is then:
%----------------------------------------------------------------------------------------
\begin{eqnarray}
d^{9}W_{123}=\left| M_{A}\right| ^{2}\frac{m_{p}}{E_{1}}\frac{d^{3}p_{1}}{%
\left( 2\pi \right) ^{3}}\frac{m_{p}}{E_{2}}\frac{d^{3}p_{2}}{\left( 2\pi
\right) ^{3}}\frac{m_{p}}{E_{3}}\frac{d^{3}p_{3}}{\left( 2\pi \right) ^{3}}, \label{eq2}
\end{eqnarray}
%----------------------------------------------------------------------------------------
where $M_{A}$ is the amplitude corresponding to the block A, i.e., accounting for the inclusive 
production of nucleons 1, 2 and 3 and other particles in the final
state $f$. To avoid cumbersome expressions 
in equations (\ref{eq1}) and (\ref{eq2}), the factors corresponding to colliding nuclei 
have been omitted since they cancel in further calculations.
Using the conventional graph technique \cite{SH68}, the expression for $M$ can 
be written in the form:
%----------------------------------------------------------------------------------------
\begin{eqnarray}
M &=&
\int \frac{d^{4}p_{1}}{\left( 2\pi \right) ^{4}}
\int \frac{d^{4}p_{2}}{\left( 2\pi \right) ^{4}}
\int \frac{d^{4}p_{3}}{\left( 2\pi \right) ^{4}} \nonumber\\
&&\frac{2m_{p}}{m_{p}^{2}-p_{1}^{2}-i0}
\frac{2m_{p}}{m_{p}^{2}-p_{2}^{2}-i0}
\frac{2m_{p}}{m_{p}^{2}-p_{3}^{2}-i0} \nonumber\\
&&i\left( 2\pi \right) ^{4}\delta^{4}\left( p_{1}+p_{2}+p_{3}-p_{t}\right) \nonumber \\
&&M_{(1,2,3\rightarrow t)} M_{A} \label{eq3}
\end{eqnarray} 
%----------------------------------------------------------------------------------------
where $M_{(1,2,3\rightarrow t)}$ is the vertex of coalescence of 1,2,3 into $t$ (proportional 
to the three-nucleon wave function in the momentum space in the nonrelativistic approximation),
$m_{p}$ the nucleon mass, the three 
fractions being the individual nucleon propagators of 1, 2 and 3. The integrals have to be 
performed over energies and momenta of the (virtual) particles. The delta functions ensure 
energy-momentum conservation at the $t$ vertex, $p_{t}=\left( \mathbf{p}_{t},E_{t}\right)$, with 
$\mathbf{p}_{t}=\mathbf{p}_{1}+\mathbf{p}_{2}+\mathbf{p}_{3}$ being the momentum of $t$, 
$E_{t}$ its energy.
The dependence of the amplitude $M_A$ on its variables (the particle momenta) is also needed 
explicitly for the calculations. In lack of a reliable theoretical form, this can be done in a 
''minimal'' way,  by using empirical shapes. The inclusive nucleon spectra usually have a 
decreasing form which can be approximated by a Gaussian function in the center of mass frame:
%----------------------------------------------------------------------------------------
\begin{eqnarray}
E_{\tiny \mbox{p}}\frac{d^3\sigma_{\tiny \mbox{p}}}{dp_{\tiny
\mbox{p}}^3} \propto \exp \left( - \mbox{\bf p}_{\tiny
\mbox{p}}^2/Q^2 \right),\label{eq4}
\end{eqnarray}
%----------------------------------------------------------------------------------------
where $Q$ defines the slope parameter of the momentum distribution. Accordingly, the
amplitude $M_A$ can be written in the following way:
%----------------------------------------------------------------------------------------
\begin{eqnarray}
M_{A} &=& C\exp \left(- \frac{ \mathbf{p}_{1}^{2}+\mathbf{p}_{2}^{2}+
\mathbf{p}_{3}^{2}}{2Q^{2}}\right) \nonumber \\
&=& C\exp \left( -\frac{\mathbf{p}_{t}^{2}}{6Q^{2}}\right) \exp \left(- \frac{\mathbf{q}^{2}}{Q^{2}}\right)
\exp \left(-\frac{3\mathbf{p}^{2}}{4Q^{2}}\right), \label{eq5}
\end{eqnarray}
%-----------------------------------------------------------------------------------------
where
%-----------------------------------------------------------------------------------------
\begin{eqnarray}
\mathbf{p}_{t} &=&\mathbf{p}_{1}+\mathbf{p}_{2}+\mathbf{p}_{3}, \nonumber\\
\mathbf{p} &=&\frac{1}{\sqrt{3}}\left( \mathbf{p}_{1}-\mathbf{p}_{2}\right), \nonumber\\
\mathbf{q} &=&\frac{1}{2\sqrt{3}}\left( \mathbf{p}_{1}+\mathbf{p}_{2}-2 \mathbf{p}_{3}\right). \label{eq6}
\end{eqnarray}
%-----------------------------------------------------------------------------------------
Assuming a statistical independence in the three nucleon production process, the inclusive 
production cross section can be written as the product of the three independent probabilities:
%-----------------------------------------------------------------------------------------
\begin{eqnarray}
\frac{d^{9}W_{123}}{dp_{1}^{3}dp_{2}^{3}dp_{3}^{3}}=\frac{1}{\sigma
_{inel}^{2}}\frac{d^{3}W_{1}}{dp_{1}^{3}}\frac{d^{3}W_{2}}{dp_{2}^{3}}\frac{%
d^{3}W_{3}}{dp_{3}^{3}},
\label{eq7}
\end{eqnarray} 
%-----------------------------------------------------------------------------------------
where $\sigma_{inel}$ is the total reaction cross-section of the colliding particles.

After integration of (\ref{eq3}), taking into account (\ref{eq5}) and (\ref{eq7}), and dividing
by the incident particle flux, the $t$ production cross section takes the form:
%-----------------------------------------------------------------------------------------
\begin{eqnarray}
E_{t}\frac{d^{3}\sigma _{t}}{dp_{t}^{3}}=\frac{96\pi ^{6}}{m_{p}^{2}\sigma
_{inel}^{2}}\left| S\right| ^{2}E_{1}\frac{d^{3}\sigma _{1}}{dp_{1}^{3}}E_{2}%
\frac{d^{3}\sigma _{2}}{dp_{2}^{3}}E_{3}\frac{d^{3}\sigma _{3}}{dp_{3}^{3}}, \label{eq8}
\end{eqnarray} 
%-----------------------------------------------------------------------------------------
with $\mathbf{p}_{1}=\mathbf{p}_{2}=\mathbf{p}_{3}$, $\mbox{\bf
p}_{\tiny \mbox{t}}=3\mathbf{p}_{1}$ and
%------------------------------------------------------------------------------------------
\begin{eqnarray}
S=\int \exp \left(- \frac{\mathbf{q}^{2}}{Q^{2}}-\frac{3}{4}\frac{\mathbf{p}^{2}}{Q^{2}%
}\right) \Psi _{t}\left( \mathbf{p},\mathbf{q}\right) \frac{d^{3}\mathbf{p}}{%
\left( 2\pi \right) ^{3}}\frac{d^{3}\mathbf{q}}{\left( 2\pi \right) ^{3}}. \label{eq9}
\end{eqnarray}
%-----------------------------------------------------------------------------------------
Where $\Psi _{t}\left( \mathbf{p},\mathbf{q}\right) \propto M_{123\rightarrow t}$ is the wave
function of the $t$ (or $^3$He) system normalized by the condition
%------------------------------------------------------------------------------------------
\begin{eqnarray}
\int \left| \Psi _{t}\left( \mathbf{p},\mathbf{q}\right) \right| ^{2}\frac{%
d^{3}\mathbf{p}}{\left( 2\pi \right) ^{3}}\frac{d^{3}\mathbf{q}}{\left( 2\pi
\right) ^{3}}=1. \label{eq10}
\end{eqnarray}
%-----------------------------------------------------------------------------------------
The factor $1/2$, accounts for nucleons and $A=3$ nuclei spins, is included in (\ref{eq8}).
The three-nucleon wave function is needed at sufficiently large momenta to compute the amplitude. 
The wave function of \cite{MU84} has been used (see appendix for discussion).

The structure of (\ref{eq8}) is the same as that of the coalescence model and the $S$ integral in 
\ref{eq8} can be straightforwardly related to the coalescence momentum:
%------------------------------------------------------------------------------------------
\begin{eqnarray}
 p_0^3 =18\sqrt{3}\pi^2 &&\int 
\frac{d^{3}\mathbf{p}}{\left(2\pi \right) ^{3}}
\frac{d^{3}\mathbf{q}}{\left( 2\pi \right) ^{3}}\nonumber\\
 &&\exp \left(- \frac{\mathbf{q}^{2}}{Q^{2}}- 
\frac{3}{4}\frac{\mathbf{p}^{2}}{Q^{2}}\right) 
\Psi _{t}\left( \mathbf{p},\mathbf{q}\right).
\label{eq11}
\end{eqnarray}
%-----------------------------------------------------------------------------------------

Thus, in the approach based on the diagram of Fig.~\ref{FEYN} and within the approximations made above, 
the coalescence momentum $p_0$ is not an adjustable parameter anymore, but it is determined by the 
inclusive proton spectrum and by the trinucleon wave function. Note that in that case, $p_0$ 
depends on the momentum distribution and should thus be energy and system dependent. 
 
%!!!!!!!!!!!!!!!!!!!!!!!!!!!!!!!!!!!!!!!!!!!!!!!!!!!!!!!!!!!!!!!!!!!!!!!!!!!!!!!!!!!!!!!!!!!!!!!!!!!!!!!!!!!!!!!!!!!!!!

\section{Application to three-antinuclei production}

In order to generalize the diagrammatic approach of the coalescence model to the production of 
$A=3$ antinuclei (noted $\overline{t}$ further below), two effects have to be taken into account: the
anisotropy of angular distributions and the threshold effects \cite{DU03a}.
%----------------------------------Fig. 2------------------------------------------------
\begin{figure}[h!]
\epsfxsize=8cm \centerline{\epsfbox{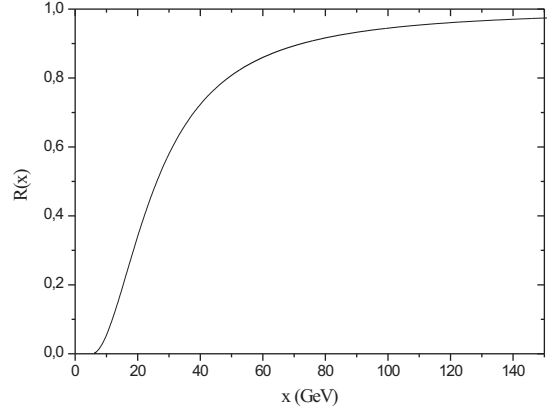}} 
\caption{ Dependence of the threshold factor $R(x)$, with $x$ as defined in the text. 
\label{THRESH}}
\end{figure}
%----------------------------------------------------------------------------------------
Isotropic angular dependence are frequently assumed in nonrelativistic collisions. However, in 
relativistic collisions, the momentum distributions are strongly anisotropic and the low
energy approximation cannot be used. To take this into account, formula (\ref{eq8}) can be easily 
generalized to any angular dependence. Assuming the inclusive nucleon production cross section 
to be given by the (parameterized) amplitude
$M_{1}\left( \mathbf{p}_{1}\right)$:
%-----------------------------------------------------------------------------------------
\begin{eqnarray}
 E_{1}\frac{d^{3}\sigma _{1}}{dp_{1}^{3}}=\left| M_{1}\left( \mathbf{p}%
_{1}\right) \right| ^{2},\label{eq12}
\end{eqnarray}
%-----------------------------------------------------------------------------------------
The cross section for $t$ production can then be written (see \ref{eq8}):
%-----------------------------------------------------------------------------------------
\begin{eqnarray}
E_{t}\frac{d^{3}\sigma _{t}}{dp_{t}^{3}} &=&
\frac{96\pi ^{6}}{m_p^2 \sigma_{inel}^2}
\left[ \int M_{1}\left( \mathbf{p}_{1}\right)
M_{2}\left( \mathbf{p}_{2}\right) M_{3}\left( \mathbf{p}_{3}\right)\right.
\nonumber\\
&&\left.\Psi _{t}\left( \mathbf{p},\mathbf{q}\right)
\frac{d^{3}\mathbf{p}}{\left(
2\pi \right) ^{3}}\frac{d^{3}\mathbf{q}}{\left( 2\pi \right) ^{3}}\right]^{2}.\label{eq13}
\end{eqnarray}
%-----------------------------------------------------------------------------------------
This model could be practically used directly to describe the production of $\overline{t}$. The 
production threshold of the antiparticle has to be taken into account however in the  
cross section calculation. The same procedure to evaluate the cross section near threshold as 
that used in \cite{DU03a} will be applied here. In nucleon-nucleon collisions, the main 
reaction producing a $\overline{t}$ particle is $NN\rightarrow \overline{\mbox{t}}+5N$.
Near the threshold of this reaction, the energy dependence of the $\overline{t}$ production cross 
section is mostly governed by the five nucleons phase space: 
$\Phi \left( \sqrt{s+m_{t}^{2}-2\sqrt{s}E_{\overline{\mbox{\tiny t}}}}%
 ; 5m_{p}\right) $,
%-------------------------------------------------------------------------------------------
\begin{eqnarray}
E_{\overline{\mbox{\tiny t}}}\frac{d^{3}\sigma _{\overline{\mbox{\tiny t}}}}{dp_{\overline{\mbox{\tiny t}}}^{3}}%
\propto \Phi \left( \sqrt{s+m_{t}^{2}-2\sqrt{s}E_{\overline{\mbox{\tiny t}}}}%
 ; 5m_{p}\right). \label{eq14}
\end{eqnarray}
%-----------------------------------------------------------------------------------------
The phase space $\Phi$ for $n$ particles with masses, momenta and energies, $m_i$, 
$\mbox{\bf p}_i$, and $E_i$ respectively, is defined in the usual way (in the center of mass)
%-----------------------------------------------------------------------------------------
\begin{eqnarray*}
&&\Phi(\sqrt{s} ; m_1, m_2, \ldots m_n) =\\
&&\prod_{i=1}^{n}  \frac{1}{(2\pi)^3} \frac{d^{3}p_i}{2E_i}
\delta^{3}\left(\sum_{i=1}^{n} \mbox{\bf p}_i\right)
\delta \left(\sum_{i=1}^{n} \mbox{\bf E}_i-\sqrt{s}\right).
\end{eqnarray*}
%-----------------------------------------------------------------------------------------
It was calculated here by using the standard CERN library program (W515, subroutine GENBOD) \cite{CERNlib}. $\sqrt{s}$ 
is the total energy of the $n$ particles in the center of mass system.

A phenomenological correction factor $R$ can thus be introduced in formulae (\ref{eq13}) which then 
reads:
%-----------------------------------------------------------------------------------------
\begin{eqnarray}
R\left( x\right)  =
\frac{\Phi \left(x; 5m_{p}\right) }{\Phi \left( x; 5 \times 0\right) }, \label{eq15}
\end{eqnarray}
%-----------------------------------------------------------------------------------------
where $x= \sqrt{s+m_{t}^{2}-2\sqrt{s}E_{\overline{\mbox{\tiny t}}}}$, and where the denominator 
contains the high energy limit of the phase space to ensure $R$ to be dimensionless and
to do not change the value of the cross section out of the space phase boundary. 
The limits of $R$ are thus:
%-----------------------------------------------------------------------------------------
\begin{eqnarray*}
&&R \rightarrow 0,\hspace{5 mm}E_{\overline{\mbox{\tiny t}}}\rightarrow E_{\overline{\mbox{\tiny t}}%
}^{\max }=\left( \frac{s+m_{t}^{2}-\left( 5m_{p}\right) ^{2}}{2\sqrt{s}}%
-m_{t}\right),  \\
%&&R \rightarrow 0,\hspace{5 mm}\sqrt{s}\rightarrow 9m_{p}, \\
&&R \rightarrow 1,\hspace{5 mm}\sqrt{s}\rightarrow \infty.
\end{eqnarray*}
%-----------------------------------------------------------------------------------------
%Furthermore, we make no assumption about mechanism of the production.

If $p_{\overline{\mbox{\tiny t}}}^2 \ll (\sqrt{s}-E_{\overline{\mbox{\tiny t}}})^2$,
the expression $\sqrt{s+m_{t}^{2}-2\sqrt{s}E_{\overline{\mbox{\tiny t}}}}$
can be replaced by $\sqrt{s}-E_{\overline{\mbox{\tiny t}}}$. This same approximation was made 
in \cite{DU03a}. The functional dependence of $R(x)$ is shown in Fig.~\ref{THRESH}.

%!!!!!!!!!!!!!!!!!!!!!!!!!!!!!!!!!!!!!!!!!!!!!!!!!!!!!!!!!!!!!!!!!!!!!!!!!!!!!!!!!!!!!!!!!!!!!!!!!!!!!!!!!!!!!!!!!!!!!!

\section{Results on antinuclei production data}

\subsection{Status of the data}
This section is introduced with a brief overview of the current experimental situation on 
the antinuclei production relevant to the present study, i.e., in proton-proton and 
proton-nucleus collisions. The antinuclei production in ion-ion collisions will be quoted 
only for completeness.

\begin{itemize}

\item As mentioned in the introduction, the experimental data on the production of mass 3 
antinuclei are extremely scarce, and much less informative than that on antideuteron production, 
with only two experiments or sets of experiments reporting on mass 3 antinuclei production in 
proton-nucleus collisions \cite{AN71,VI74,BO78,BU80}. Note that there are no 
experimental data available on the production of these antinuclei in proton-proton collisions. 
The production of $\overline{^3\mbox{He}}$ has been observed recently in various heavy ion studies like $Pb+Pb$ 
collisions at ultra relativistic incident energies \cite{NA52}. These data are out of the scope 
of the present work. They will not be discussed here (see \cite{DU03a} for a discussion).

\item Coalescence calculations require the antiproton production cross section to be known
for antiproton momenta equal to approximately one third of the $A=3$ antinuclei momenta.
Unfortunately, in most experiments the differential cross sections for antiproton and
$A=3$ antinuclei productions were not measured at this momentum.  
The $\overline{p}$ cross section thus had to be extrapolated to the appropriate kinematical region when 
no other data were available, which of course, introduces additional uncertainty in the 
calculations.

\end{itemize}

The three nucleon wave functions needed in the calculations are much less well know than 
the deuteron wave function. In addition, the same wave function will be used for $^{3}He$ 
and $t$ nuclei (see appendix). The inaccuracy on the tri-nucleon wave functions is thus 
another source of uncertainty.

The total reaction cross-section used in the calculations was described by means of the
parameterization proposed in \cite{LE83}. 
%Similar results are obtained with \cite{LE83}.

\subsection{Proton Aluminium collision data at 70GeV/$c$}
The antinuclei produced in the Serpukhov experiments \cite{AN71,VI74} were 
obtained from a 70~GeV/$c$ proton beam incident on an aluminium target at 27~ mrad scattering 
angles and 20~GeV/$c$ for $\overline{^3\mbox{He}}$, and 0° and 25 GeV/$c$ for $\overline{t}$. The inclusive 
antiproton cross sections were available from \cite{BI69} and \cite{BU70} in the same 
kinematical conditions. 

\begin{figure}[h!]
\epsfxsize=9cm \centerline{\epsfbox{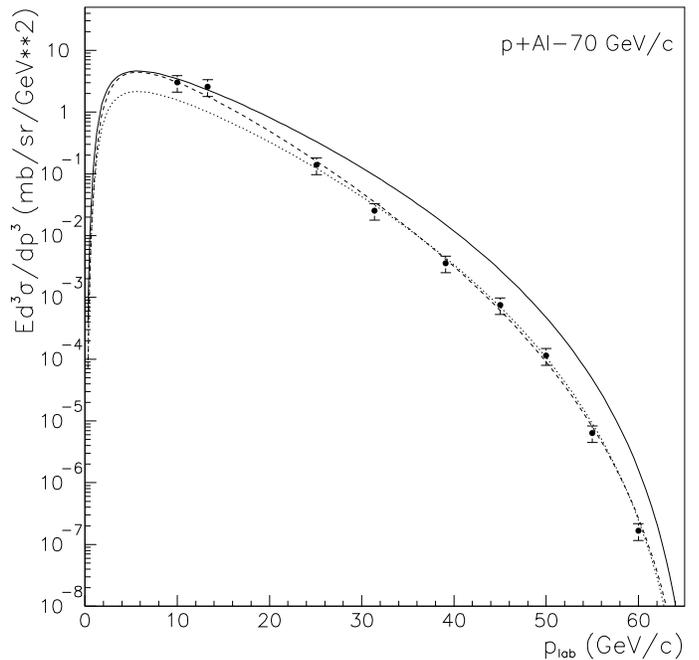}} 
%{Fig3.eps}} 
\caption{Inclusive differential cross section for antiproton production in
p+Al collisions as a function of the total momentum in the laboratory
frame from \cite{AN71}, compared to calculated values as discussed in the
text. \label{PBARS1}}
\end{figure}

In Fig.~\ref{PBARS1} the $\overline{p}$ cross section data from \cite{AN71} are compared with the results 
of fits using a functional form \cite{DU03b}. The solid curve corresponds to a fit of a large 
sample of $p+A\rightarrow\overline{p}$ data from 12 up to 400 GeV incident energies not including those 
from reference \cite{AN71} which were found not to be compatible with the other sets of data 
\cite{DU03b}.
%The same conclusion was reached in \cite{***}.
The calculated values are in fair agreement 
with the two lowest momentum data points (which were obtained by extrapolation from measurements at other 
angles). They overestimate the other data points by a factor of 2 to 4. The dotted curve is a 
renormalization of the solid curve by a factor $\approx 2.5$ to fit these latter points, while the dashed 
curve corresponds to the fit of the single set of data points shown on the figure, which parameters 
however give quite poor agreement with the other sets of data \cite{DU03b}.

%---------------------------Fig .3------------------------------------------------
%---------------------------------------------------------------------------------
It must be emphasized that the low momentum region, say $p_{lab}<10$ GeV/$c$, which is the useful 
region for the coalescence calculations, with $\mathbf{p}_{\overline{p}}\approx\mathbf{p}_
{\overline{\mbox{\tiny t}}}/3$ is particularly important here, with unfortunately no data 
point from direct measurement available over the relevant range.
%--------------------------Fig. 4-------------------------------------------------
\begin{figure}[h!]
\epsfxsize=9cm \centerline{\epsfbox{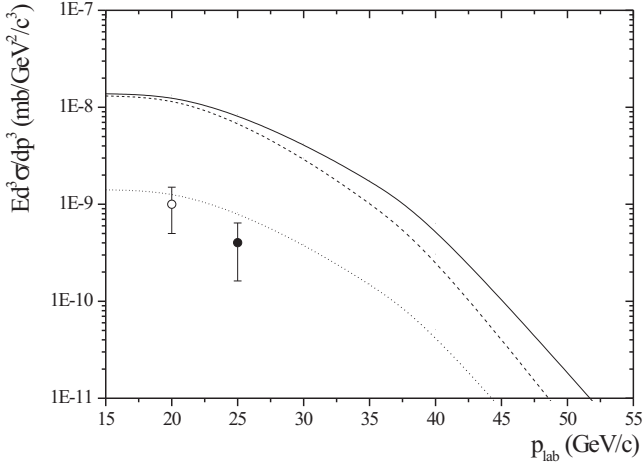}} 
%Fig4.eps}} 
\caption{The
inclusive differential cross-section for \={t} \cite{AN71} (open circles)
and $\overline{^{3}\mbox{He}}$ \cite{VI74} (black circles) production by 70 GeV  
protons on Al target as a function of the total momentum in the laboratory frame 
compared to calculations using the microscopic model of coalescence and the same three 
different parameterizations of the antiproton production cross-section as shown on figure 
\ref{PBARS1}, with the same graphical conventions. \label{SERPUK}}
\end{figure}
%---------------------------------------------------------------------------------

Fig.~\ref{SERPUK} compares the calculations for the $A=3$ production cross section for the 
three parameterizations shown on Fig.~\ref{PBARS1} with the experimental data. The 
calculations using the global fit renormalized to the 70 GeV/$c$ data (see Fig.~\ref{PBARS1})
give by far the best agreement with the $\overline{t}$ data (dotted curve). The other two sets 
of $\overline{p}$ cross section parameters overestimate the data by a sound order of magnitude. 
This is apparently consistent with the larger $\overline{p}$ cross section predicted by these two 
sets of parameters for low $\overline{p}$  momentum region to which the $\overline{t}$
cross section is most sensitive. The factor of about 2 between the $\overline{p}$  cross sections 
predicted by the two groups of parameters translates into a factor of about 10 for the 
$\overline{t}$  cross section because of the approximately cubic dependence of the latter on the 
$\overline{p}$  cross section. However it is somewhat puzzling that this agreement is obtained 
with parameters which are not consistent with the whole body of $\overline{p}$  data \cite{DU03b}. 

\subsection{Proton beryllium collision data at 200 GeV/$c$}
In the CERN experiments \cite{BO78,BU80}, $\overline{p}$, $\overline{t}$  and $\overline{^{3}\mbox{He}}$ 
were produced in proton-beryllium collisions at 200, 210, and 240 GeV/$c$ and detected at 0 
degree scattering angle \cite{BU80}, while $\overline{p}$  were measured at 200 GeV/$c$ \cite{BO78} on the
same targets. 
For these data however, the %$\overline{p}$, $\overline{t}$ and $\overline{^{3}\mbox{He}}$ 
production cross sections were measured as the ratios to the $\pi ^{-}$ production cross 
sections at the same momentum. The knowledge of the corresponding experimental $\pi ^{-}$ 
production cross section, or a good parameterisation of the latter, is thus required in 
order to allow the values of the $\overline{p}$, $\overline{t}$ and $\overline{^{3}\mbox{He}}$ 
production cross sections to be calculated.

\begin{figure}[h!]
\epsfxsize8cm \centerline{\epsfbox{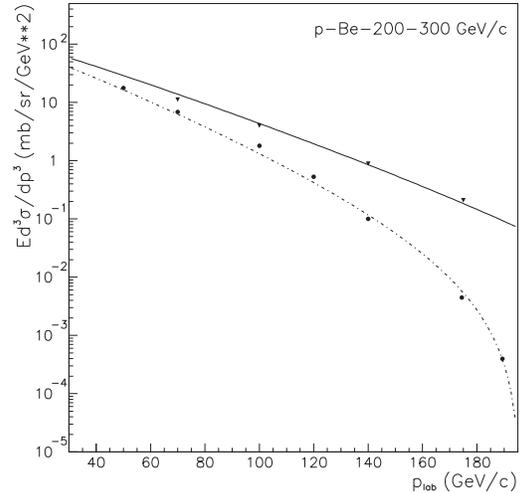}} \caption{Experimental inclusive 
differential cross-section for $\pi ^{-}$ production in $p+Be$ collisions \cite{BA74} 
(symbols) at 200 GeV/$c$ (full circles) and 300 GeV/$c$ (full triangles) compared with the 
functional form \ref{FPI}. \label{BAPI}}
\end{figure}

\begin{figure}[h!]
\epsfxsize=8cm \centerline{\epsfbox{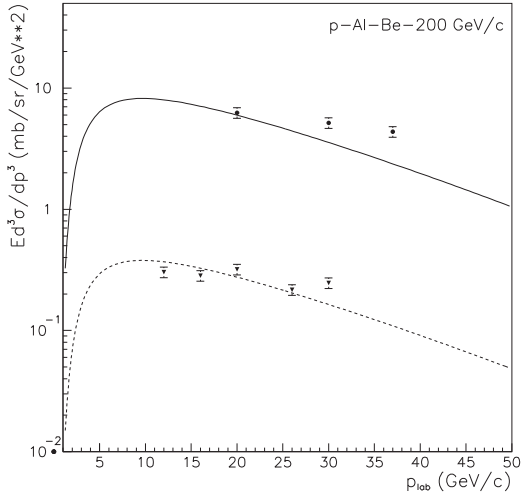}} \caption{Experimental inclusive 
differential cross-sections for $\overline{p}$ production in $p+Al$ collision (circles),
and in $p+Be$ collision ($\times 10^{-1}$, triangles), [18] compared with the results of fits
using a functional form (curves) \cite{DU03b}. \label{FPBAR} }                                                                                                                                                                                                                                                                                                                                                                                                                                                                                                                                                                                                                                                                                                                                                                                                                                                                                                                                                                                                                                                                                                                                                                                                                                                                                                                                                                                                                                                                                                                                                                                                                                                                                                                                                                                                                                                                                                                                                                                                                                                                                                                                                                                                                                                                                                                                                                                                                                                                                                                                                                                                                                                                                                                                                                                                                                                                                                                                                                                                                                                                                                                                                                                                                                                                                                                                                                                                                                                                                                                                                                                                                                                                                                                                                                                                                                                                                                                                                                                                                                                                                                                                                                                                                                                                                                                                                                                                                                                                                                                                                                                                                                                                                                                                                                                                                                                                                                                                                                                                                                                                                                                                                                                                                                                                                                                                                                                                                                                                                                                                                                                                                                                                                                                                                                                                                                                                                                                                                                                                                                                                                                                                                                                                                                                                                                                                                                                                                                                                                                                                                                                                                                                                                                                                                                                                                                                                                                                                                                                                                                                                                                                                                                                                                                                                                                                                                                                                                                                                                                                                                                                                                                                                                                                                                                                                                                                                                                                                                                                                                                                                                                                                                                                                                                                                                                                                                                                                                                                                                                                                                                                                                                                                                                                                                                                                                                                                                                                                                                                                                                                                                                                                                                                                                                                                                                                                                                                                                                                                                                                                                                                                                                                                                                                                                                                                                                                                                                                                                                                                                                                                                                                                                                                                                                                                                                                                                                                                                                                                                                                                                                                                                                                                                                                                                                                                                                                                                                                                                                                                                                                                                                                                                                                                                                                                                                                                                                                                                                                                                                                                                                                                                                                                                                                                                                                                                                                                                                                                                                                                                                                                                                                                                                                                                                                                                                                                                                                                                                                                                                                                                                                                                                                                                                                                                                                                                                                                                                                                                                                                                                                                                                                                                                                                                                                                                                                                                                                                                                                                                                                                                                                                                                                                                                                                                                                                                                                                                                                                                                                                                                                                                                                                                                                                                                                                                                                                                                                                                                                                                                                                                                                                                                                                                                                                                                                                                                                                                                                                                                                                                                                                                                                                                                                                                                                                                                                                                                                                                                                                                                                                                                                                                                                                                                                                                                                                                                                                                                                                                                                                                                                                                               
\end{figure}

Fortunately, the $p+Be\rightarrow\pi^{-}+X$ cross section has been measured at 200 and 300 
GeV/$c$ incident momentum in \cite{BA74} in similar kinematical conditions as in the CERN 
experiment.
The measured distributions have been fit by means of the following functional form, inspired 
from ref \cite{KALI}:
\begin{eqnarray}
E\frac{d^{3}\sigma }{dp^{3}}\left(\pi ^{-}\right) =C_{1}\sigma _{in}\left(
1-x\right) ^{C_{2}}e^{-C_{3}x}e^{-C_{5}p_{\perp}},
\label{FPI}
\end{eqnarray}
where $x=E ^{-}/E ^{*}_{max}$ ($E ^{*}$  is the total energy of the inclusive particle in
the center of mass frame, $\sigma _{in}$ is the total reaction cross section for the 
system in collision, $\sqrt{s}$ is the total energy of the system and $p_{\perp}$ the 
transverse momentum of the emitted particle. The values of the parameters obtained are 
given in Table \ref{PARPI} and the results of this fit are presented in Fig. \ref{BAPI}.
%--------------------------Fig. 5-------------------------------------------------
%---------------------------------------------------------------------------------
The parameterization (\ref{FPI}) has been used to extract the experimental $\overline{p}$ 
production cross sections \cite{BO78}. The resulting cross section values are compared 
in Fig. \ref{FPBAR} with the results of the fit of a functional form to a large sample of 
$p+A\rightarrow\overline{p}+X$ data from 12 GeV/$c$ up to 400 GeV/$c$ 
incident momenta \cite{DU03b}. It is seen that the data points derived previously and the 
calculated values are in fair agreement. This consistency gives confidence to the following 
steps of the analysis for the evaluation of the $\overline{t}$ and $\overline{^{3}\mbox{He}}$ 
production cross sections.
%--------------------------table. 2-----------------------------------------------
\begin{table}
\begin{center}
\begin{tabular}{ccccc}\hline
parameter & $C_{1}$ & $C_{2}$ & $C_{3}$ & $C_{4}$   \\
value     &   0.94  &  1.88   &    7.05  &   1.69 \\   \hline \\
\end{tabular}
\caption{\small Values of the parameters of relation \ref{FPI} obtained by fitting the 
$\pi ^{-}$ production cross sections for 200 and 300 GeV/$c$ protons on Beryllium. 
\label{PARPI}}
\end{center}
\end{table}
%---------------------------------------------------------------------------------
%--------------------------Fig. 6-------------------------------------------------
%---------------------------------------------------------------------------------
%--------------------------Fig. 7-------------------------------------------------
\begin{figure}[h!]
\epsfxsize=7cm \centerline{\epsfbox{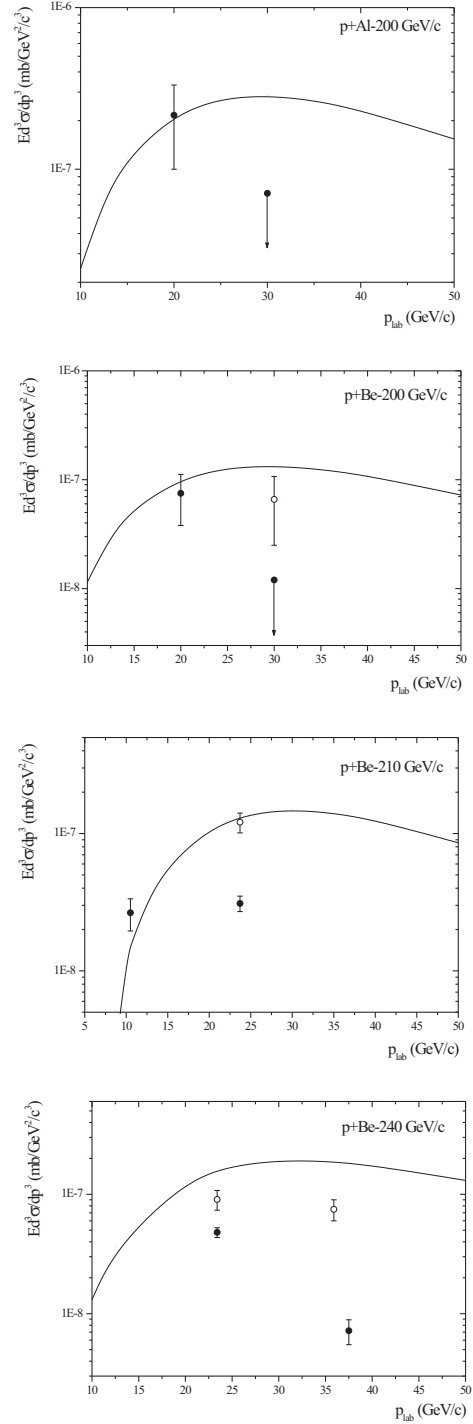}} 
\caption{Experimental inclusive differential cross-section for $\overline{t}$ (full circles) 
and $\overline{^{3}\mbox{He}}$ (open circles) production in $p+Al$ and $p+Be$ collisions, 
compared to calculations using the microscopic model of coalescence (solid line).
\label{THEB}}
\end{figure}
%---------------------------------------------------------------------------------
In Fig. \ref{THEB}, the $\overline{t}$ and $\overline{^{3}\mbox{He}}$ production cross 
sections are compared to calculations using the microscopic model of coalescence.
The agreement between experimental and calculated values varies from poor to good. On the
average however data and calculations are within one order of magnitude. This result should be 
considered as a success in account of the numerous sources of uncertainties of the 
calculations and of the limited accuracy of the measurements. Note also that refs \cite{BO78} 
and \cite{BU80} report experimental values in disagreement by a factor of 2. Furthermore
the $\overline{t}$ and $\overline{^{3}\mbox{He}}$ production cross section, measured at the same momentum,
should be in principly close to each other (this fact is clearly seen in the same experiment for $t$ 
and $^{3}\mbox{He}$ production) whereas, in this experiment, they are quite different.
 
%!!!!!!!!!!!!!!!!!!!!!!!!!!!!!!!!!!!!!!!!!!!!!!!!!!!!!!!!!!!!!!!!!!!!!!!!!!!!!!!!!!!!!!!!!!!!!!!!!!!!!!!!!!!!!!!!!!!!!!
\section{Conclusion}
It has been shown in this work that the diagrammatic approach to the coalescence model 
developed previously previously can successfully account for the mass 3 antinuclei
production cross section in proton-nucleus collisions over wide kinematical conditions without any additional parameter.
These calculations require a good knowledge of the antiproton production cross-section and 
of the three-nucleon wave function. These results would be further used to calculate
$\overline{t}$ and $\overline{^{3}\mbox{He}}$ flux in cosmic rays.

%!!!!!!!!!!!!!!!!!!!!!!!!!!!!!!!!!!!!!!!!!!!!!!!!!!!!!!!!!!!!!!!!!!!!!!!!!!!!!!!!!!!!!!!!!!!!!!!!!!!!!!!!!!!!!!!!!!!!!!
\section*{Appendix}
In this appendix, we briefly remind how the wave functions of
the trinucleon is written in \cite{MU84}, while in this paper slightly
different definitions have been used. A useful analytical parameterization of the
bound trinucleon wave function is obtained from solving the
Faddeev equation with the Reid soft-core potential.

The total wave function of the triton is written as a sum of 3 Faddeev components
%-----------------------------------------------------------------------------------------
\begin{eqnarray*}
\Psi =\sum_{i=1}^{3} \Psi _{t}^{i}\left( \mathbf{q}_{i},\mathbf{p}_{i}\right).
\end{eqnarray*}
%-----------------------------------------------------------------------------------------
In (\ref{eq9})-(\ref{eq13}), we make use of only one Faddeev component $\Psi _{t}$,
while due the exchange symmetry of two-nucleons in the triton, all
the three Faddeev components are identical. Furthermore, Faddeev
components $\Psi _{t}$  are decomposed in terms of their partial
wave components with respect to the spin-isospin and angular
momentum basis $\phi_{\alpha }(\mathbf{\hat{p}},\mathbf{\hat{q}})$
.
%-----------------------------------------------------------------------------------------
\begin{eqnarray*}
\Psi _{t}(\mathbf{p},\mathbf{q})=\sum_{\alpha }\psi _{\alpha }(p,q)\phi
_{\alpha }(\mathbf{\hat{p}},\mathbf{\hat{q}}),
\end{eqnarray*}
%-----------------------------------------------------------------------------------------
with, if \textbf{p}$_{i}$ ($i=1,2,3$) is the nucleon momenta
%-----------------------------------------------------------------------------------------
\begin{eqnarray*}
\mathbf{p} =\frac{1}{2}\left( \mathbf{p}_{1}-\mathbf{p}_{2}\right),
\mathbf{q} =\frac{1}{2\sqrt{3}}\left( \mathbf{p}_{1}+\mathbf{p}_{2}-2%
\mathbf{p}_{3}\right).
\end{eqnarray*}
%-----------------------------------------------------------------------------------------
The following normalization is used
%-----------------------------------------------------------------------------------------
\begin{eqnarray*}
\int \left| \Psi _{t}\left( \mathbf{p},\mathbf{q}\right) \right| ^{2}d^{3}%
\mathbf{p}d^{3}\mathbf{q} =
\sum_{\alpha } \int dpdqp^{2}q^{2}\left| \psi _{\alpha }\left( p,q\right) \right| ^{2}
=1.
\end{eqnarray*}
%-----------------------------------------------------------------------------------------
Note that, in these expressions, the definition of $\mathbf{p}$ and the normalization
differ from (\ref{eq6}) and (\ref{eq10}). $\Psi _{t}$ is a sum of the partial wave
state $\alpha$ which is a label for the following physical
quantities:
\begin{itemize}
\item $L$, the angular momentum of the pair of nucleons (1-2).
\item $l$, the angular momentum of nucleon 3 according to the center of the mass
of the pair of nucleons (1-2).
\item $\mathcal{L}$, the total angular momentum of the triton.
\item $s$, the spin of the pair of nucleons (1-2).
\item $\mathcal{S}$, the total spin of the triton.
\item $T$, the isospin of the pair of nucleons (1-2).
\end{itemize}

Only two components label $\alpha$ were taken into account, $\alpha=1,2$.
\begin{itemize}
\item For $\alpha=1$, $L=l=\mathcal{L}=0$, $s=1$, $\mathcal{S}=1/2$ and $T=0$.
\item For $\alpha=2$, $L=l=\mathcal{L}=s=0$, $\mathcal{S}=1/2$ and $T=1$.
\end{itemize}
Of course, the fact to consider only two partial wave state is an
approximation which gives the probability of 89.25\% of trinucleon
being in the partial wave state $\alpha$. In \cite{MU84}, the
parameterization for $\psi _{\alpha }\left( p,q\right)$ is given by
%-----------------------------------------------------------------------------------------
\begin{eqnarray*}
\psi _{\alpha }\left( p,q\right)&&=
p^{L}p^{l}\left( p^{2}+\Omega _{p1}^{2}\right) ^{-1}
\prod_{m=1}^{3}\left( q^{2}+\Omega _{qm}^{2}\right) ^{-1}\\
&&\sum_{i=1}^{6}\sum_{j=1}^{6}\frac{C_{ij}}{\left( p^{2}+\mu
_{i}^{2}\right) \left( q^{2}+\nu _{j}^{2}\right) },
\end{eqnarray*}
%-----------------------------------------------------------------------------------------
with $\Omega _{p1}$, $\Omega _{qm}$, $\mu_{i}$, $\nu _{j}$ and $C _{ij}$  
all depending on the partial wave label $\alpha$. The
numerical values of these coefficients can be found in
\cite{MU84}.

$^{3}He(ppn)$ and $t(pnn)$ are considered to have the same wave
function. Altough, because of the presence of the Coulomb
interaction, these two wave functions are slightly different,
this difference is negligible compared to the other uncertainties of
present calculations.

%!!!!!!!!!!!!!!!!!!!!!!!!!!!!!!!!!!!!!!!!!!!!!!!!!!!!!!!!!!!!!!!!!!!!!!!!!!!!!!!!!!!!!!!!!!!!!!!!!!!!!!!!!!!!!!!!!!!!!!

%!!!!!!!!!!!!!!!!!!!!!!!!!!!!!!!!!!!!!!!!!!!!!!!!!!!!!!!!!!!!!!!!!!!!!!!!!!!!!!!!!!!!!!!!!!!!!!!!!!!!!!!!!!!!!!!!!!!!!!

\end{document}